# Downlink Interference Estimation without Feedback for Heterogeneous Network Interference Avoidance

Siyi Wang, Weisi Guo[†], Mark D. McDonnell
Institute for Telecommunications Research, University of South Australia, Australia
[†]School of Engineering, The University of Warwick, United Kingdom
siyi.wang@mymail.unisa.edu.au, [†]weisi.guo@warwick.ac.uk, mark.mcdonnell@unisa.edu.au

***Abstract*—In this paper, we present a novel method for a base station (BS) to estimate the total downlink interference power received by any given mobile receiver, without information feedback from the user or information exchange between neighbouring BSs. The prediction method is deterministic and can be computed rapidly. This is achieved by first abstracting the cellular network into a mathematical model, and then inferring the interference power received at any location based on the power spectrum measurements taken at the observing BS. The analysis expands the methodology to a $\mathsf{K}$-tier heterogeneous network and demonstrates the accuracy of the technique for a variety of sampling densities. The paper demonstrates the methodology by applying it to an opportunistic transmission technique that avoids transmissions to channels which are overwhelmed by interference. The simulation results show that the proposed technique performs closely or better than existing interference avoidance techniques that require information exchange, and yields a 30% throughput improvement over baseline configurations.**

## I. Introduction

Due to increased urbanisation, the growing density of wireless network nodes in cities is causing the performance of links to be increasingly interference-limited, as opposed to propagation-limited in the past. In developed cities, the density of co-frequency cellular macro base stations (macro-BSs) has reached over 7 per km$^2$, and Wi-Fi access-points has reached over 700 per km$^2$. As cellular networks look to expand their capacity via *spectrum reuse*, the number of cellular low-powered-nodes (e.g., femto-cells and relay-nodes) is set to grow rapidly to 12 million by 2014. One of the main drawbacks of additional cells is the excess interference, which is hard to predict due to the high resolution required for small cell planning [1]. The success of capacity growth by will depend on research techniques that can effectively mitigate the co-channel interference [2]–[4].

Existing literature over the past few years has shown that avoiding co-channel interference in the networks of high intensity interference can improve the long-term system throughput [4]–[6]. However, to coordinate interference avoidance on the radio-resource-management (RRM) level, there typically needs to be a large volume of coordination information synchronised between multiple BSs[1]. That is to say, for an OFDMA system, each BS needs to know whether neighbouring BSs are transmitting on each radio resource. This level of coordination is taxing on the *backhaul capacity* and any *delay* in information sharing can cause the scheme's performance to falter. Simplifying coordination has been demonstrated to be effective, whereby coordination is limited to adjacent dominant interfering BSs [7], but in reality a larger cluster of BSs needs to cooperate in order for a more flexible architecture and more effective interference reduction.

This paper proposes a technique for interference estimation that does not require information sharing between BSs or user equipments (UEs). This is achieved by estimating the transmissions of neighbouring BSs, and inferring the resulting interference power at any point of interest. That is to say, we propose to infer the average channel quality at *any random point*, based on some passive measurements taken at a single *observational point* [8]. In the first part of the paper, we resent the novel method for channel quality estimation using stochastic geometry. We then expand the framework to consider an open access co-channel $\mathsf{K}$-tier heterogeneous network. In the second part, we demonstrate the proposed technique with respect to opportunistic interference reduction, which is shown to approach the accuracy of information exchange on the X2 interface.

## II. Formulation

### A. Stochastic Geometry Model of Interference Power

In this paper we consider an OFDMA based multiple-access system, which is utilised in 3GPP Long-Term-Evolution (LTE), IEEE 802.16 WiMax and IEEE 802.20 WiMAN. In particular, we consider the performance of downlink (DL) channels, that are interference-limited (AWGN is negligible). Fig. 1 illustrates an example LTE network. At any particular time and radio-resource-block (RRB) snap-shot, a certain set of BSs are transmitting (shaded) and a certain set of BSs are silent. We now derive an analytical expression for the *mean aggregated interference power* in the DL channel of an arbitrarily located UE. We employ the spatial Poisson point process (SPPP) denoted as $\Phi$ [9]–[11]. Within the network, a fixed density of $\chi$ cells (e.g., macro-BSs) are deployed, of which at any time instance and on any resource block, $\lambda \leqslant \chi$ cells are actively transmitting.

The paper denotes random variables (R.V.) as capital letters and their particular values as lower-case letters. We define a random point in space that is of range $r$ to the nearest

[1]coordination information is sent via channels such as the X2 channel in 4G LTE

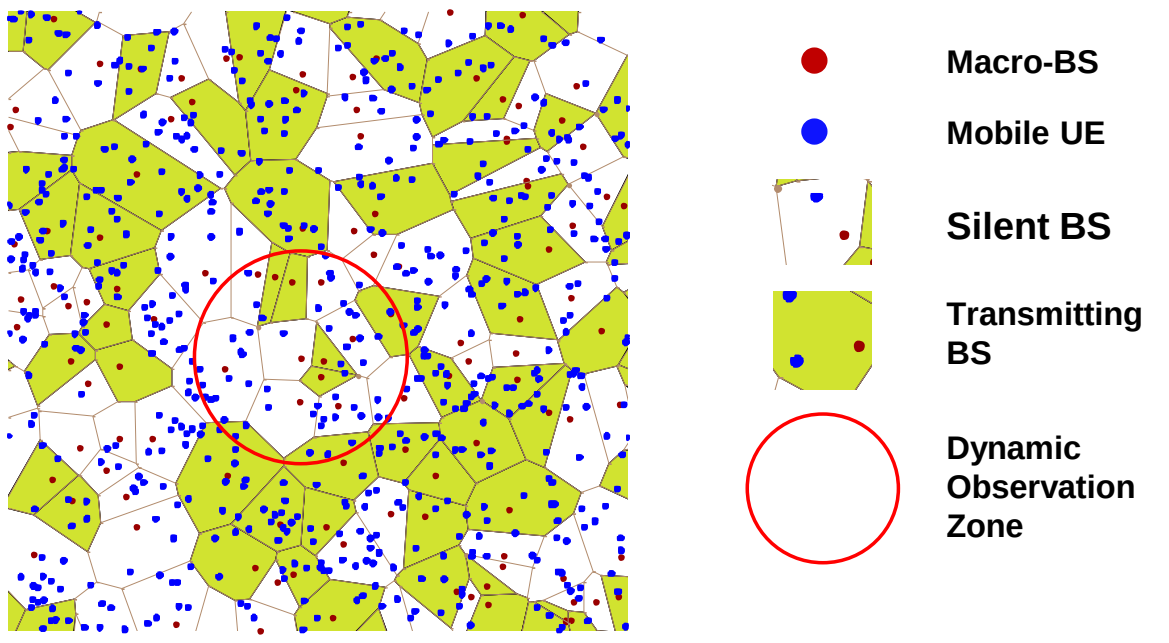

Fig. 1. Illustration of a 4G LTE-A cellular network with interference avoidance implemented at the macro-BSs.

serving cell (index $l$). The total DL interference received at any particular point by an arbitrarily located UE is $I_r$:

$$I_r = \sum_{\substack{i \in \Phi \\ i \neq l}} h_i P \Lambda r_i^{-\alpha}, \tag{1}$$

where $h_i$ is the multi-path fading gain from an interference BS (index $i$), $r_i$ is the distance from interfering transmitter to receiver, $P$ is the transmit power, $\Lambda$ is the pathloss constant and $\alpha$ is the pathloss distance exponent [12]. For urban environments, the paper considers Rayleigh fading with mean 1 and defines R.V. $G = HP\Lambda$, which therefore follows the exponential distribution with mean $\beta$ denoted by $G \sim \exp(\beta)$, where $\beta = 1/P\Lambda$. The symbols used in this paper and their assumed values can be found in Table I.

The moment generating function of $I_R$ is:

$$\begin{aligned} M_{I_R}(-s;\alpha) &= \mathscr{L}[f_{I_R}(I_r)](s) = \mathbb{E}(e^{-sI_r}), \\ &= \mathbb{E}_\Phi\left\{\mathbb{E}_G\left[\exp\left(-s\sum_{\substack{i\in\Phi\\ i\neq l}} g_i r_i^{-\alpha}\right)\right]\right\}, \\ &= \mathbb{E}_\Phi\left\{\prod_{\substack{i\in\Phi\\ i\neq l}} \mathbb{E}_G\left[\exp\left(-s g_i r_i^{-\alpha}\right)\right]\right\}, \\ &= \mathbb{E}_\Phi\left(\prod_{\substack{i\in\Phi\\ i\neq l}} \frac{\beta}{\beta + s r_i^{-\alpha}}\right). \end{aligned} \tag{2}$$

Further derivation of the expressions for the moment generating function can be found in *Appendix A*. The probability density function (PDF) of $I_R$ is obtained on taking the inverse Laplace transform:

$$\begin{aligned} f_{I_R}(I_r;\alpha) &= \mathscr{L}^{-1}[M_{I_R}(-s)](I_r), \\ &= \mathscr{L}^{-1}\left\{\exp\left[-\pi\lambda\sqrt{\frac{s}{\beta}}Q(r,\mathsf{R},4)\right]\right\}(I_r), \\ &= \frac{\sqrt{\pi}\lambda Q(r,\mathsf{R},4)}{2I_r^{3/2}}\exp\left\{-\frac{[\pi\lambda Q(r,\mathsf{R},4)]^2}{4\beta I_r}\right\}, \end{aligned} \tag{3}$$

where $Q(r,\mathsf{R},4) = \arctan(\mathsf{R}) - \arctan(r)$. The cumulative distribution function (CDF) of $I_R$ is thus given by:

$$f_{I_R}(\zeta;4) = \int_0^\zeta f_{I_R}(I_r;4)\,\mathrm{d}I_r = \operatorname{erfc}\left[\frac{\pi\lambda Q(r,\mathsf{R},4)}{2\sqrt{\beta\zeta}}\right]. \tag{4}$$

The average interference power is:

$$\mathbb{E}(I_r) \approx \left[\frac{\pi\lambda Q(r,\mathsf{R},4)}{2\sqrt{\beta}\operatorname{erfc}^{-1}(0.5)}\right]^2, \tag{5}$$

where the median is used for the purpose of this paper to show key trends and results. The mean can be explicitly found, as shown in Eq. (20) in *Appendix B*, but it is quite convoluted.

The paper now expands this result to a heterogeneous network with K-tiers of different cells (i.e., macro, micro, pico, and femto). Each tier has an associated active cell density $\lambda_\mathsf{k}$ and transmit power $P_\mathsf{k}$. Each UE is attached to the cell with the strongest received signal power and receives interference from all other cells in every co-channel tier. The proof can be found in *Appendix B*. The average interference power can also be found using Eq. (19):

$$\mathbb{E}(I_{r_\mathsf{i}}) \approx \left[\frac{\pi Q(r_\mathsf{il},R,4)}{2\operatorname{erfc}^{-1}(0.5)}\sum_{\mathsf{k}=1}^{\mathsf{K}}\frac{\lambda_\mathsf{k}}{\sqrt{\beta_\mathsf{k}}}\right]^2, \tag{6}$$

for a total of K tiers. For $\mathsf{K} = 1$, Eq. (6) reduces to Eq. (5).

### B. Dynamic Interference Observation Zone

In order to realistically and accurately represent the interference power from a number of co-channel transmissions from a large set of BSs, we need to define the an *observation zone*. As shown in Fig. 2, an observation zone is a circular area that is centred on the observing BS and covers a number of BSs, such that the aggregated interference power from them is sufficiently representative.

Using simulation results, the paper first performs a preliminary analysis into the number of DL interference BSs that needs to be considered in order to obtain an accurate interference power value. Whilst this analysis has been performed before for a traditional hexagonal grid model [13], it deserves to be tested for a more realistic non-hexagonal cell distribution model, such as a realistic cellular network deployment [14]. Fig. 3 shows the total DL interference power received at various points in a cell, that is subjected to interference from other cells constrained by distance (expressed as a factor of the cell coverage radius). First we observe that the total interference power is a monotonically increasing function with the observation zone radius R. This is apparent by examining either Fig. 3 or Eq. (6), which is a quadratic with respect to R. It is also apparent that the interference observation zone size R is related to the location of the UE, which is defined by the distance to the serving cell ($r$). Intuitively, the results show that for cell-edge UEs (large $r$), considering nearby interferers is sufficiently accurate, whereas for cell-centre UEs (small $r$), considering a larger set of interference sources is needed. Therefore, in order to establish a reasonable zone of control, we define the zone to have a radius R m, which sufficiently accurately portrays the interference environment for a particular UE. Therefore, for a reasonably accurate representation of the interference power, we define:

$$\frac{\mathrm{d}\mathbb{E}(I_R)}{\mathrm{dR}} \leqslant \varphi, \tag{7}$$

where $\varphi$ is a gradient threshold used to control the accuracy of the estimate. The greater the interference observation zone, the smaller the value of $\varphi$ achieved, as shown in Fig. 3. A unique value of R can be found numerically with ease with a gradient descent or similar search algorithm.

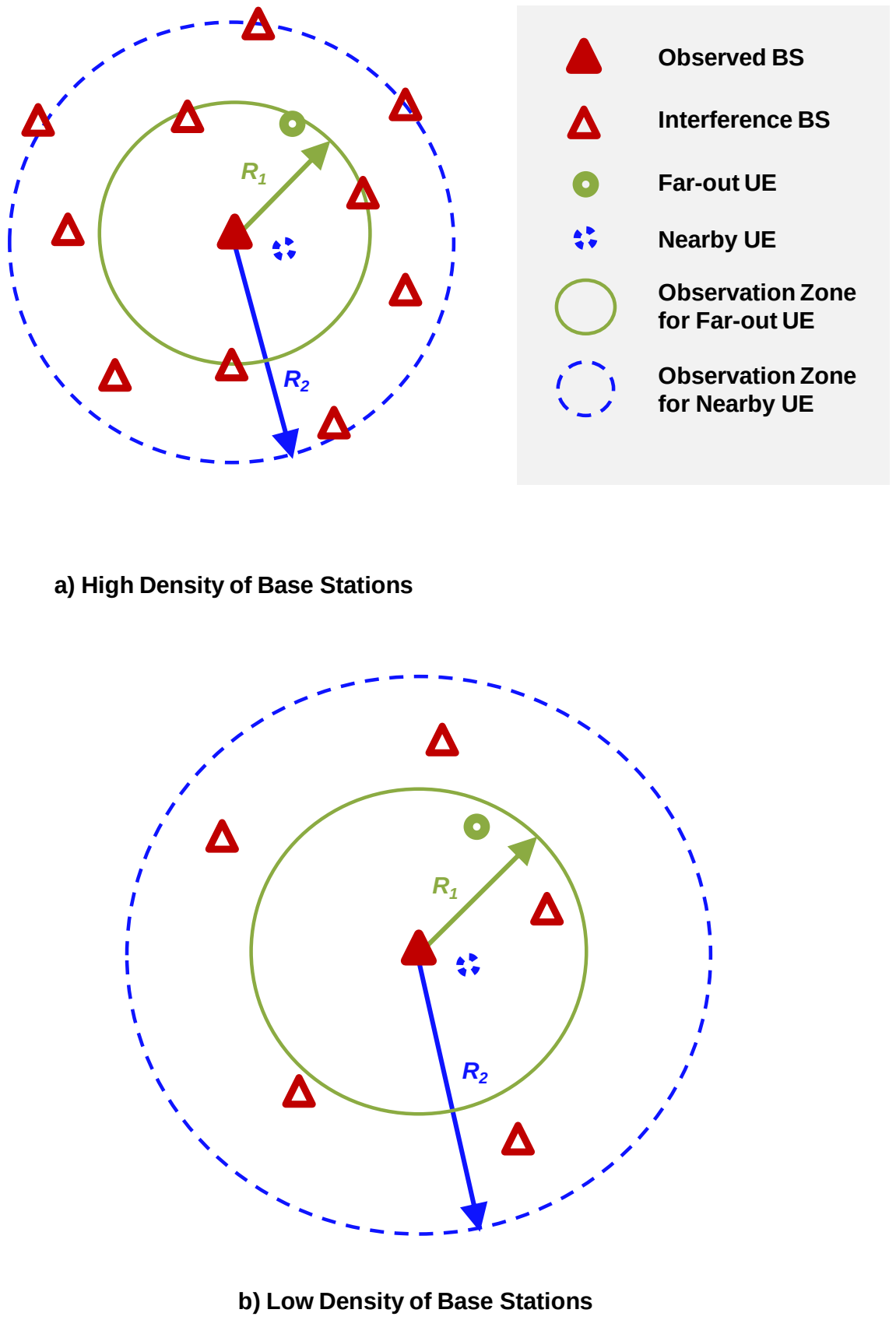

Fig. 2. Illustration of dynamic observation zone for different UE positions $r$, under different deployed cell densities $\lambda$.

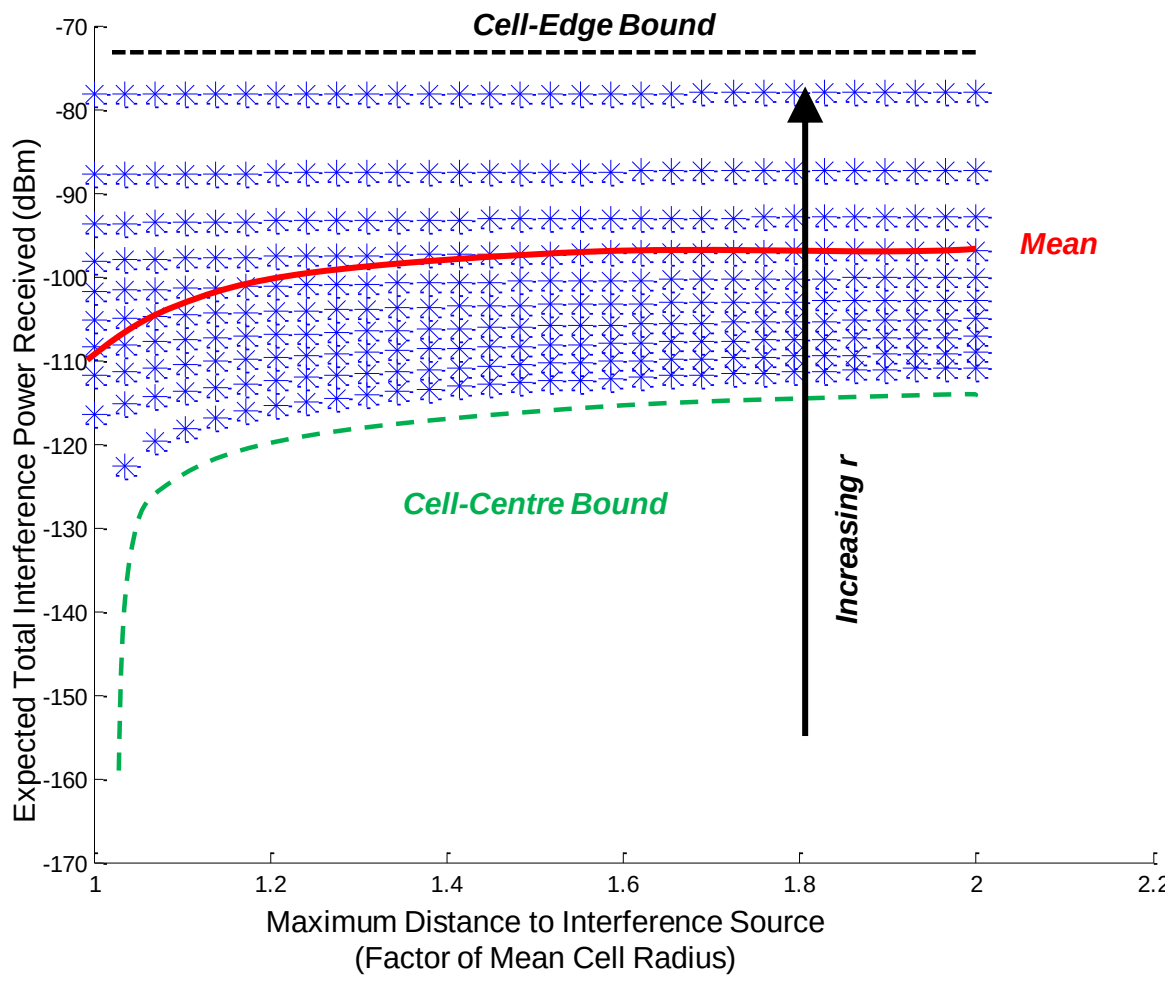

Fig. 3. Plot of total DL interference power received as a function of the observation zone radius R, for different UE-BS distances $r$.

As illustrated in Fig. 2, we can also observe several *intuitive* but useful relationships on the size of the observation zone by examining Eq. (5):

- *Cell Density*: as cell density increase ($\lambda \uparrow$), the observation zone needed to sufficiently portray total interference power decreases (R $\downarrow$).
- *UE Location*: as the UE distance to serving observation BS increases ($r \uparrow$), the observation zone needed decreases (R $\downarrow$).
- *Transmit Power*: as the BS transmit power increases ($P \uparrow$, $\beta \propto 1/P$), the observation zone needed to sufficiently portray total interference power decreases (R $\downarrow$).

In summary, each BS will employ a dynamic observation zone of radius R, where the size of the zone varies with the targeted UE's location $r$, the actively transmitting cell density $\lambda$, and the cell transmit power $P$. The paper now examines how to estimate the average interference power at any UE (location $r$), based on measurements taken at its observational serving-BS.

## III. Interference Power Estimation

So far, the paper has proposed theoretical methods for determining the number of interference cells that one should consider to accurately portray the interference power. We have reinforced our argument of a dynamic observation zone with simulation results. The paper now shows how one can estimate the interference power at a UE, without any channel feedback or information exchange between any BSs.

We assume interference power measurements are taken at each BS by using a simple power spectrum analyser. In terms of the mathematical framework, we measure the average interference power spectrum at $r = d$, where $d$ is small ($\sim$ metres). In reality, the power spectrum analyser is placed at the null of the serving-BS antenna pattern or shielded from its radiation. Let us define this measurement distance to be $r = d$ away from the BS. Therefore, from Eq. (5), the *estimated* active interference cell density $\lambda'$ can be expressed as a function of the average measured interference power on a particular sub-band at any particular time instance. For a homogeneous (1 tier) network, this is explicit, but only the aggregated densities can be expressed for a heterogeneous network:

$$\begin{cases} \lambda' = \frac{2\sqrt{\beta\mathbb{E}(I_{d,\lambda})}\mathrm{erfc}^{-1}(0.5)}{\pi Q(d,\mathrm{R}_d,4)} & \text{homogeneous} \\ \lambda'_\mathrm{K} = \frac{2\sqrt{\mathbb{E}(I_{d,\lambda})}\mathrm{erfc}^{-1}(0.5)}{\pi Q(d,\mathrm{R}_d,4)} & \text{heterogeneous} \end{cases}, \tag{8}$$

where $\lambda'_\mathrm{K} = \sum_{\mathrm{k}=1}^{\mathrm{K}} \frac{\lambda'_\mathrm{k}}{\sqrt{\beta_\mathrm{k}}}$.

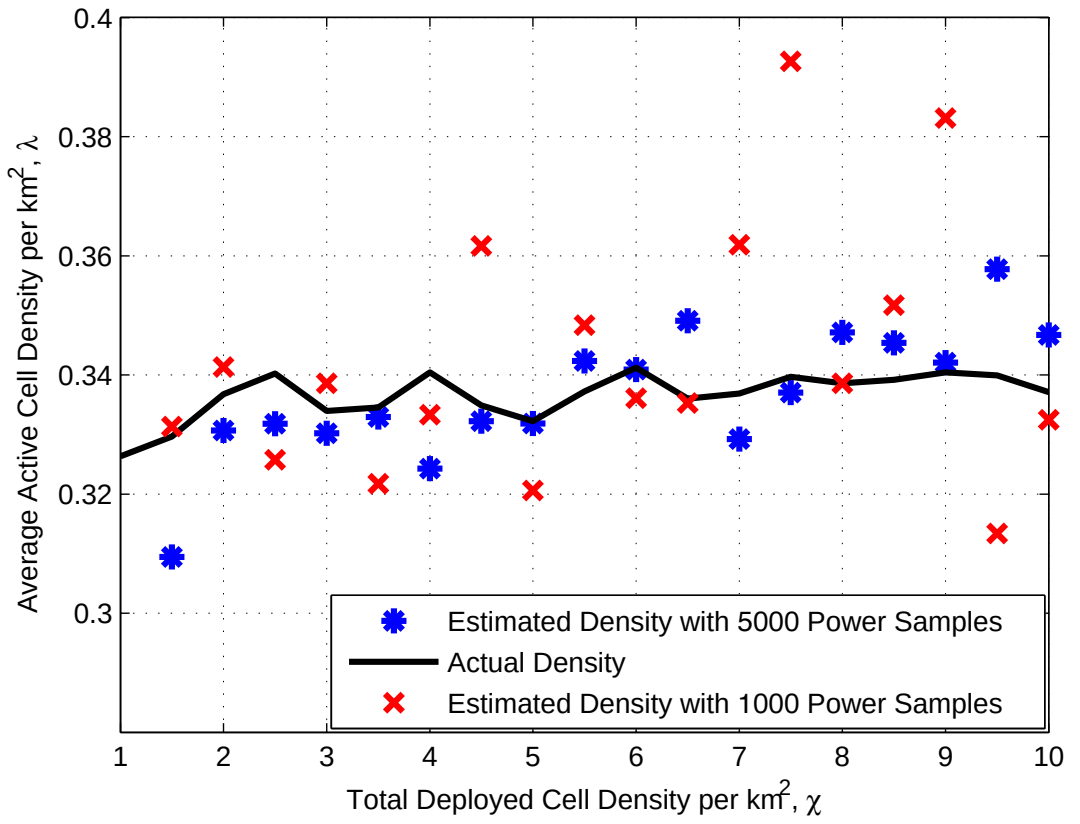

Fig. 4. Validation plot of simulated actively transmitting BS density ($\lambda$) and inferred BS density ($\lambda'$), as a function of shifting total deployed cell density ($\chi$).

TABLE I
SYSTEM PARAMETERS

| *Parameter* | *Symbol and Value* |
|---|---|
| Cellular Network | 4G LTE OFDMA DL |
| Bandwidth | 20 MHz |
| Number of Cells | 35 |
| PHY Layer | MCS SISO [12] |
| Deployed BS Density | $\chi$, 5 per km$^2$ |
| Active BS Density | $\lambda$ |
| BS Transmit Power | $P$, 40 W |
| Avg. Number of UEs | 10 per BS |
| Fading Gain | $h$, Rayleigh |
| Pathloss Model | Urban Micro [12] |
| BS to UE Distance | $r$ |
| Total Interference Power | $I_r$ |
| Observation Zone Radius | R |
| Interference Measurement Point | $d$ |
| Throughput SIR Threshold | $\epsilon$, $-6$ dB |

To validate our inference method, the results in Fig. 4 show that the simulated actively transmitting BS density ($\lambda$) and inferred BS density ($\lambda'$), as a function of shifting total deployed cell density ($\chi$). The inferred density arrives from Eq. (8) and can be used to calculate the interference power at any point $r$ using Eq. (9). The results show that an interference power sampling rate of 5000 per quasi-static period of traffic will yield an accuracy of 94%, and 1000 samples will yield an accuracy of 85%. Therefore, a high number of samples at the BS's measurement point $d$ is needed in order to obtain accurate results.

Given the estimated active interference cell density $\lambda'$ at a known and fixed detection point $r = d$, the resulting real-time (zero delay) interference power at any point $r$ can be found by using Eq. (5) by substituting in the estimated variable $\lambda'$:

$$\boxed{\mathbb{E}(I_{r,\lambda}) \approx \mathbb{E}(I_{d,\lambda}) \left[\frac{Q(r, \mathsf{R}_r, 4)}{Q(d, \mathsf{R}_d, 4)}\right]^2}, \quad (9)$$

where the measurement point $d$ and the observation zone radii $R$ are deterministic and known. It is worth noting that in order to use the same statistical pathloss model at the measurement point and for all the UEs, the paper measures the interference at the base of each BS (i.e., ground level). Therefore the value of $d$ is in fact the antenna height of the BS.

An interesting and important result is that the method for estimating interference power is the same for a homogeneous and heterogeneous network. In the Appendix, we show that Eq. (9) is the same for both a 1-tier homogeneous and an open-access K-tier heterogeneous network. The authors suspect this is only true for the median average properties of both network configurations. In fact, the Appendix shows how the mean in Eq. (20) is more complicated for a K-tier heterogeneous network. However, as stated earlier, we use the median average in this paper to demonstrate the application of this technique.

Armed with the ability to estimate the *real-time* interference power without receiving a delayed feedback from UEs or exchanging transmission information between BSs, the paper now employs the estimated interference power to implement interference avoidance.

## IV. INTERFERENCE AVOIDANCE RESULTS

In this section, we consider an OFDMA based cellular network, with a physical-layer given by discrete modulation-and-coding schemes (MCSs) [12]. There exists an important SIR lower-bound ($\epsilon$), below which packet switched throughput is strictly zero.

The interference avoidance mechanism employed in this paper allows each BS to make a binary downlink transmission decision process to a UE, based on the interference power it infers that the UE will receive. The inference is based on the previously discussed technique, given by Eq. (9). The transmission policy is to avoid transmission (and interference to other users) when the resulting throughput is estimated to be lower than the minimum target SIR $\epsilon$. Therefore, the transmission power for serving-BS to each UE is:

$$P = \begin{cases} P & \text{for: } \frac{r^{-4}/\beta}{N+\mathbb{E}(I_{r,\lambda})} > \epsilon \\ 0 & \text{otherwise} \end{cases}, \quad (10)$$

where the UE's SIR at a range $r$ is given by $\frac{r^{-4}/\beta}{N+\mathbb{E}(I_{r,\lambda})}$ for an AWGN power of $N$ on a particular RRB.

The simulation results presented in Fig. 5 show an implementation of the proposed interference avoidance scheme, in comparison with a baseline hard frequency reuse 1 setup, and a Sequential Game Coordinated (SGC) interference avoidance scheme [7]. The simulation parameters can be found in Table I.

*1) Low Traffc Regime:* At a *low cell activity level*, which corresponds to a low traffic load in the area, the overall interference power received at any point is low. The achievable maximum throughput aggregated across all DL links in the observed cell is high for all schemes. For the baseline hard frequency reuse 1 (HFR1) scheme, the maximum capacity is 58 Mbit/s per cell. For the two interference avoidance schemes, namely the proposed and SGC scheme, the maximum capacity is 77 and 82 Mbit/s per cell, respectively. The 30% improvement is primarily due to the improved interference avoidance which is relatively easy to accomplish at low traffic loads in the SGC scheme [7]. The proposed performance is slightly lower because radio resource re-allocation is not performed.

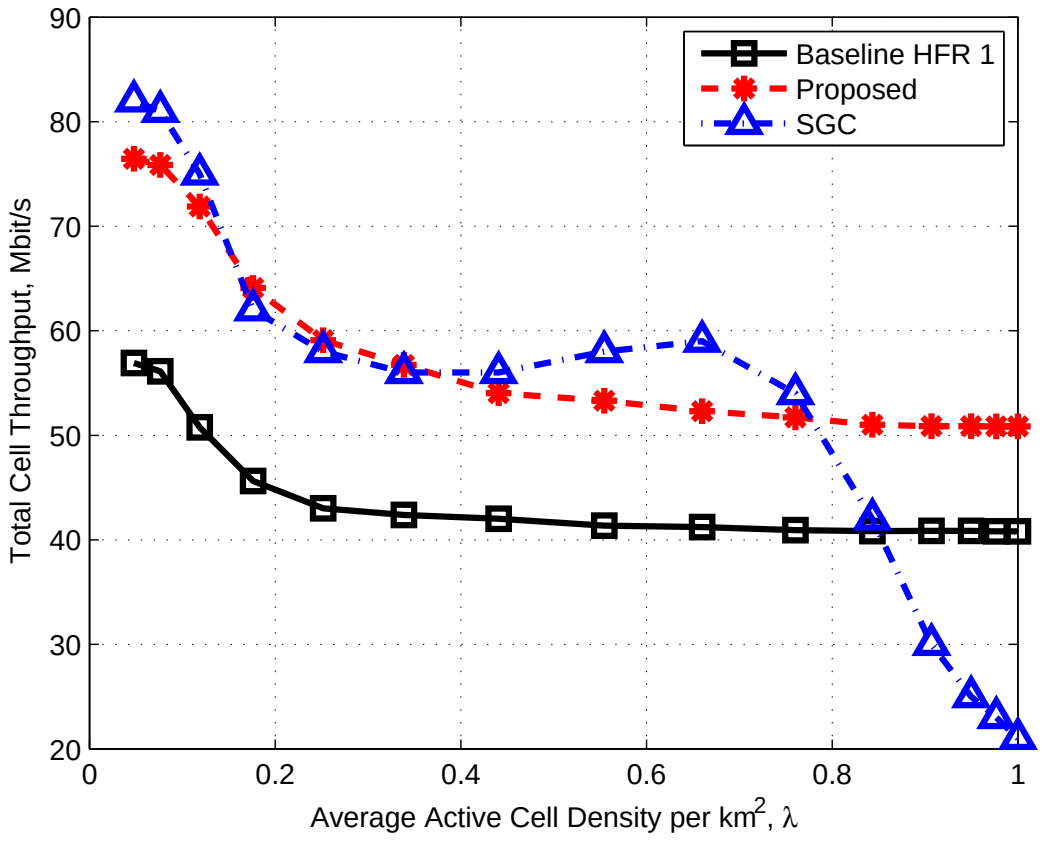

Fig. 5. Plot of simulated maximum cell throughput as a function of the neighbouring active transmit cell density $\lambda$.

*2) Medium to High Traffc Regime:* At a *medium to high cell activity level*, which corresponds to a high traffic load in the area, the overall interference power received at any point is much higher. For the baseline HFR1 scheme, the maximum capacity falls and converges to 42 Mbit/s per cell. For the SGC scheme, the throughput falls dramatically, recovers, then falls to 20 Mbit/s. The reason for the general decrease is that it increasingly operates as a HFR 2 scheme, where one cell's resources are off and another cell's are on. This reduces the interference at the cost of the bandwidth efficiency. The dramatic initial dip at $\lambda = 0.3$ is due to the coordination complexities of the scheme, which are detailed in [7]. For the proposed scheme, TDD cell switching is not operated, and the cell throughput saturates at 50 Mbit/s, which is a 19% improvement over the standard HFR 1 scheme, recommended and operated by most 3G HSPA and 4G LTE networks.

## V. Conclusions

In this paper, we presented a novel method for base stations to estimate the total downlink interference power to any given mobile receiver, without any channel feedback from the user or information exchange between BSs. The prediction method is deterministic and can be computed rapidly. This is achieved by first abstracting the cellular network into a stochastic geometry mathematical model. The interference power received by any user at arbitrary location (point $r$) can be inferred based on a single power spectrum measurement taken at the base of the serving-BS. Furthermore, we expand the framework to a K-tier heterogeneous network and show that the results are equally applicable. The method is applied to opportunistic interference avoidance and the results demonstrate the accuracy of this technique for a variety of network configurations. Monte-Carlo simulation results show that the proposed technique performs closely or better than existing interference avoidance techniques that require information exchange, and yields a 30% throughput improvement over baseline configurations.

## Acknowledgement

The work in this paper is partly supported by the Australian Research Council (ARC), and the EPSRC Urban Science Doctoral Training Centre at the University of Warwick.

## Appendix A
## Interference Power at Arbitrary Position

Given the moment generating function $M_{I_R}(-s;\alpha)$ in Eq. (2), the next step follows from the i.i.d. distribution of $G$ and its further independence from the SPPP $\Phi$:

$$M_{I_R}(-s;\alpha) = e^{-2\pi\lambda \int_0^{+\infty}\left(1-\frac{\beta}{\beta+sv^{-\alpha}}\right)v\ \mathrm{d}v}. \tag{11}$$

Since the Laplace transform of a function $f(t)$, defined for all real numbers $t \geqslant 0$, is the function $F(s)$, given by:

$$F(s) \triangleq \mathscr{L}[f(t)](s) = \int_0^{+\infty} e^{-st} f(t)\ \mathrm{d}t. \tag{12}$$

Unlike existing work, which considers an infinite number of cells, we need to consider interference up to a realistic distance of the observation zone radius $\mathsf{R}$. Eq. (11) evaluated at $-s$ can be re-written as the Laplace transform of the PDF of $I_R$:

$$\begin{aligned} M_{I_R}(-s;\alpha) &= \mathscr{L}[f_{I_R}(I_r)](s) = \mathbb{E}(e^{-sI_r}), \\ &= \exp\left[-\pi\lambda\left(\frac{s}{\beta}\right)^{\frac{2}{\alpha}} Q(r,\mathsf{R},\alpha)\right], \end{aligned} \tag{13}$$

where $Q(r,\mathsf{R},\alpha) = \int_r^{\mathsf{R}} \frac{1}{1+u^{\frac{\alpha}{2}}}\ \mathrm{d}u$. For $\alpha = 4$, $M_{I_R}(-s;4)$ can be expressed as

$$M_{I_R}(-s;4) = \exp\left[-\pi\lambda\sqrt{\frac{s}{\beta}}Q(r,\mathsf{R},4)\right], \tag{14}$$

where $Q(r,\mathsf{R},4) = \arctan(\mathsf{R}) - \arctan(r)$. Indeed, alternative pathloss exponents can also yield expressions, some of which can be found in [10].

## Appendix B
## K-tier HetNet Interference Power

Let us consider a collection of cells that belong to different transmission classes (macro, micro, pico, and femto). They are modelled by SPPP $\Phi_{\mathsf{i}}$ of intensity $\lambda_{\mathsf{i}}$ ($\mathsf{i} = 1 \ldots \mathsf{K}$) in the Euclidean plane, respectively. Then, a heterogeneous cellular deployment can be modelled as a K-tier network where each tier models the cells of a particular class and the K SPPPs are assumed to be spatially independent in location. The mobile

UEs are also arranged according to some independent SPPP $\Phi_{\mathsf{u}}$ of intensity $\lambda_{\mathsf{u}}$. Without loss of generality, the analysis of the model is focused on a typical UE located at the origin. The downlink received SIR assuming the user connects to $\mathsf{l}^{\text{th}}$ BS in an $\mathsf{i}^{\text{th}}$-tier is calculated as below ignoring antenna gain ($G$) and log-normal shadowing ($\mathcal{S}$):

$$\gamma_{\mathsf{il}} = \frac{P_{\mathsf{i}}\Lambda h_{\mathsf{il}} r_{\mathsf{il}}^{-\alpha}}{\sum_{\mathsf{k}=1}^{\mathsf{K}} \sum_{\substack{\mathsf{j}\in\Phi_{\mathsf{k}} \\ \backslash \mathrm{BS}_{\mathsf{il}}}} P_{\mathsf{tk}}\Lambda h_{\mathsf{kj}} r_{\mathsf{kj}}^{-\alpha}} = \frac{P_{\mathsf{i}}\Lambda h_{\mathsf{il}} r_{\mathsf{il}}^{-\alpha}}{I_{\mathsf{kj}}}, \quad (15)$$

where $r_{\mathsf{il}}$ and $r_{\mathsf{kj}}$ are the distance between the typical mobile user and its associated home BS and the $\mathsf{j}^{\text{th}}$ interfering BSs in the $\mathsf{k}^{\text{th}}$-tier, respectively. $h_{\mathsf{il}}$ and $h_{\mathsf{kj}}$ follow the defined exponential distribution. $P_{\mathsf{i}}$ is the transmit power of BSs in the $\mathsf{i}^{\text{th}}$-tier. Define the new random variable $\mathsf{G}_{\mathsf{i}} = P_{\mathsf{i}}\Lambda \mathsf{H}_{\mathsf{i}}$ and hence $\mathsf{G}_{\mathsf{i}} \sim \exp(\beta_{\mathsf{i}})$ where $\beta_{\mathsf{i}} = \frac{1}{P_{\mathsf{i}}\Lambda}$.

For *open access* defined as a typical user can connect to the BS in any tier, the moment generating function of $\mathfrak{I}_{\mathsf{kj}}$ is defined as:

$$\begin{aligned} M_{\mathfrak{I}_{\mathsf{kj}}}(-s;\alpha) &= \mathscr{L}[f_{\mathfrak{I}_{\mathsf{kj}}}(I_{\mathsf{kj}})](I_{\mathsf{kj}}) = \mathbb{E}(e^{-sI_{\mathsf{kj}}}), \\ &= \mathbb{E}_{\Phi_{\mathsf{k}}}\left\{ \mathbb{E}_{\mathsf{G}_{\mathsf{k}}}\left[ \exp\left( -s \sum_{\mathsf{k}=1}^{\mathsf{K}} \sum_{\substack{\mathsf{j}\in\Phi_{\mathsf{K}} \\ \backslash \mathrm{BS}_{\mathsf{il}}}} P_{\mathsf{tk}}\Lambda h_{\mathsf{kj}} r_{\mathsf{kj}}^{-\alpha} \right) \right] \right\}, \\ &= \prod_{\mathsf{k}=1}^{\mathsf{K}} \exp\left[ -\pi\lambda_{\mathsf{k}} \left( \frac{s}{\beta_{\mathsf{k}}} \right)^{\frac{2}{\alpha}} Q(r_{\mathsf{il}}, \mathsf{R}, \alpha) \right]. \end{aligned} \quad (16)$$

For a special case where $\alpha = 4$, $M_{\mathfrak{I}_{\mathsf{kj}}}(-s;4)$ reduces to:

$$M_{\mathfrak{I}_{\mathsf{kj}}}(-s;4) = \prod_{\mathsf{k}=1}^{\mathsf{K}} \exp\left[ -\pi\lambda_{\mathsf{k}} \sqrt{\frac{s}{\beta_{\mathsf{k}}}} Q(r_{\mathsf{il}}, \mathsf{R}, 4) \right]. \quad (17)$$

Therefore, the associated PDF and CDF of the aggregate interference are express as:

$$\begin{aligned} & f_{\mathfrak{I}_{\mathsf{kj}}}(I_{\mathsf{kj}};4) = \mathscr{L}^{-1}[M_{\mathfrak{I}_{\mathsf{kj}}}(-s;4)](I_{\mathsf{kj}}) \\ &= \frac{\sqrt{\pi} Q(r_{\mathsf{il}}, \mathsf{R}, 4)}{2 I_{\mathsf{kj}}^{3/2}} \sum_{\mathsf{k}=1}^{\mathsf{K}} \frac{\lambda_{\mathsf{k}}}{\sqrt{\beta_{\mathsf{k}}}} \exp\left\{ -\frac{[\pi Q(r_{\mathsf{il}}, \mathsf{R}, 4)]^2}{I_{\mathsf{kj}}} \sum_{\mathsf{k}=1}^{\mathsf{K}} \frac{\lambda_{\mathsf{k}}^2}{4\beta_{\mathsf{k}}} \right\}, \end{aligned} \quad (18)$$

and

$$F_{\mathfrak{I}_{\mathsf{kj}}}(I_{\mathsf{kj}};4) = \operatorname{erfc}\left[ \frac{\pi Q(r_{\mathsf{il}}, \mathsf{R}, 4)}{2\sqrt{I_{\mathsf{kj}}}} \sum_{\mathsf{k}=1}^{\mathsf{K}} \frac{\lambda_{\mathsf{k}}}{\sqrt{\beta_{\mathsf{k}}}} \right], \quad (19)$$

respectively.

The expected value of the aggregated interference received at the typical user end is given by:

$$\begin{aligned} \mathbb{E}(\mathfrak{I}_{\mathsf{kj}})(\mathsf{R};4) &= \int_0^{\mathsf{R}} \omega f_{\mathfrak{I}_{\mathsf{kj}}}(\omega;4)\, \mathrm{d}\omega = \sqrt{\pi \mathsf{R}} Q(\mathsf{r}_{\mathsf{il}}, \mathsf{R}, 4) \sum_{\mathsf{k}=1}^{\mathsf{K}} \frac{\lambda_{\mathsf{k}}}{\sqrt{\beta_{\mathsf{k}}}} \\ &\times \exp\left\{ -\frac{[\pi Q(\mathsf{r}_{\mathsf{il}}, \mathsf{R}, 4)]^2}{4\mathsf{R}} \left( \sum_{\mathsf{k}=1}^{\mathsf{K}} \frac{\lambda_{\mathsf{k}}}{\sqrt{\beta_{\mathsf{k}}}} \right)^2 \right\} \\ &- \frac{[\pi Q(\mathsf{r}_{\mathsf{il}}, \mathsf{R}, 4)]^2}{2} \left( \sum_{\mathsf{k}=1}^{\mathsf{K}} \frac{\lambda_{\mathsf{k}}}{\sqrt{\beta_{\mathsf{k}}}} \right)^2 \operatorname{erfc}\left( \frac{\pi Q(\mathsf{r}_{\mathsf{il}}, \mathsf{R}, 4)}{2\sqrt{\mathsf{R}}} \sum_{\mathsf{k}=1}^{\mathsf{K}} \frac{\lambda_{\mathsf{k}}}{\sqrt{\beta_{\mathsf{k}}}} \right), \end{aligned} \quad (20)$$